\documentclass[fleqn,usenatbib]{mnras}
\usepackage{graphicx}
\usepackage{amsmath,amssymb,amsfonts}
\usepackage[cal=cm]{mathalfa}
\DeclareRobustCommand{\VAN}[3]{#2}
\let\VANthebibliography\thebibliography
\def\thebibliography{\DeclareRobustCommand{\VAN}[3]{##3}\VANthebibliography}
\newcommand{\be}{\begin{equation}}
\newcommand{\ee}{\end{equation}}
\newcommand{\daa}{\Delta\alpha/\alpha}

\title[Convergence properties and varying $\alpha$]{Convergence properties of fine structure constant measurements using quasar absorption systems}

\author[Webb, Lee]{John K. Webb$^{1,2,3}$,
Chung-Chi Lee$^3$.
\\\\
$^1$Institute of Astronomy, University of Cambridge, Madingley Road, Cambridge, CB3 0HA, UK.\\
$^2$Clare Hall, University of Cambridge, Herschel Road, Cambridge CB3 9AL, UK.\\
$^3$Big Questions Institute, Level 4, 55 Holt Street, Surry Hills, Sydney, NSW 2010, Australia.}

\date{Accepted XXX. Received YYY; in original form ZZZ}
\pubyear{2023}

\begin{document}
\label{firstpage}
\pagerange{\pageref{firstpage}--\pageref{lastpage}}
\maketitle

\begin{abstract}
Searches for spacetime variations of fundamental constants have entered an era of unprecedented precision. New, high quality quasar spectra require increasingly refined analytic methods. In this article, a continuation in a series to establish robust and unbiased methodologies, we explore how convergence criteria in non-linear least squares optimisation impact on quasar absorption system measurements of the fine structure constant $\alpha$. Given previous claims for high-precision constraints, we critically examine the veracity of a so-called ``blinding'' approach, in which $\alpha$ is fixed at the terrestrial value during the model building process, releasing it as a free parameter only after the ``final'' absorption system kinematic structure has been obtained. We show that this approach results in an extended flat canyon in $\chi^2$-$\alpha$ space, such that convergence is unlikely to be reached, even after as many as 1000 iterations. The fix is straightforward: $\alpha$ must be treated as a free parameter from the earliest possible stages of absorption system model building. The implication of the results presented here is that all previous measurements that have used initially-fixed $\alpha$ should be reworked.
\end{abstract}


\section{Introduction} \label{sec:intro}

Searches for spacetime variations of fundamental constants are motivated by theoretical expectations, hints in some previous empirical results, and improved technologies, facilities and procedures. Of particular importance is new instrumentation such as the ESPRESSO spectrograph on the VLT \citep{espresso2021} and new laser frequency comb calibration facilities \citep{Steinmetz2008, DinkoPhD, Fortier2019, Milakovic2020, Probst2020, Zhao2021}, as well as the forthcoming high resolution spectrograph ANDES on the ELT \citep{Marconi2016, ELT2018}.

To maximise return from these new and forthcoming facilities, it is important to seek improvements, where possible, amongst all aspects of our current analytic procedures \citep{Webb2022}. In a series of recent papers we have scrutinised existing methods, identified shortcomings, and have developed more advanced numerical and theoretical procedures. That work includes the application of AI methods to spectral modelling \citep{gvpfit2017, Bainbridge2017, Lee2020AI-VPFIT}. Recently, strong bias in some existing quasar absorption measurements has been revealed, \citep{Webb2021, Lee2021, Webb2022, Lee2023bias}.

Our recent work has focused on measurements of the fine structure constant $\alpha$ in quasar absorption systems, parameterised using $\daa = (\alpha_z - \alpha_0)/\alpha_0$, where the subscripts $z, 0$ indicate redshift and the terrestrial value and where the fine structure constant $\alpha = e^2/4\pi\epsilon_0\hbar c$ in SI units. Our measurements have made use of the non-linear least squares code, {\sc vpfit} \citep{ascl:VPFIT2014, web:VPFIT}. The vast majority of existing quasar absorption $\alpha$ measurements in the literature have used this code. The {\sc vpfit} theoretical background and recent enhancements are described in \cite{WebbVPFIT2021, Lee2021Addendum}. 

One aspect of {\sc vpfit} that had not previously been studied in any detail concerns convergence, but a number of quite different non-linear least squares applications motivate doing so. \cite{Forbes2009} suggest that Gauss-Newton (GN) can sometimes exhibit slow convergence and minimise the objective function at each iteration of the process, progressing the GN direction and the negative gradient. Forbes and Bartholomew-Biggs apply their modified GN method to five modelling problems and whilst they find evidence for improvement, they also find that it is difficult to identify a generalised method for convergence improvement, noting that specific fine-tuning of their method is probably required for any particular highly nonlinear problem.

\cite{Transtrum2012} suggest that slow convergence and robustness to initial guesses are complimentary problems and that methods for improving convergence speed can decrease robustness to initial guesses, and {\it vice versa}. Levenberg-Marquardt (LM) algorithms can also be painfully slow to converge when iterations proceed along a narrow canyon (which commonly exists for problems involving a large number of parameters), and suggest several ways in which convergence can be accelerated. One consequence of a flat $\chi^2$-parameter space is that algorithms may sometimes push parameters to non-physical (even infinite) values. \cite{Transtrum2012} provide an interesting discussion in which they make a distinction between {\it convergence criteria} and {\it stopping criteria}, and propose a ``geodesic acceleration'' correction to the LM step, by including second order corrections in the Taylor approximation of the residuals. 

\cite{Fratarcangeli2020} report similar lengthy convergence issues to those we report in this paper. The context of their investigation is image reconstruction. They derive significant speed gains by using LM to locally linearise the problem, decomposing the linear problem into small blocks, using the local Schur complement, obtaining a more compact linear system without loss of information, generality, or accuracy. The other main gain from their work is that the system becomes parallelised and scalable, making it suitable for use with graphics processing units. A process such as this could be explored for {\sc vpfit}.

In \cite{Lee2023bias} (L23 hereon), we showed that if $\daa$ is held fixed at zero whilst developing the absorption system model, then subsequently allowed to vary, the final measurement remains locked in a local $\chi^2$ minimum near the fixed value, so is likely to bear little or no similarity to the statistically correct solution. In the present paper, we develop the L23 analysis further, investigating the impact of non-linear least squares convergence using {\sc vpfit}. Stopping criteria must be carefully considered for $\daa$ measurements. Previous best-fit quasar absorption system models have adopted a ``standard'' practice in terminating iterations; when the change in the goodness of fit between two successive iterations, $\Delta\chi^2$, falls below some ``reasonable'' threshold, iterations are stopped and the ``best-fit'' model obtained. Previous measurements have {\it assumed} that final parameter values have been reached in this way. However, if we force algorithms like {\sc vpfit} to continue iterating well beyond what would normally be considered as a ``reasonable'' stopping criterion, could the $\daa$ solution slowly drift from the previous ``final'' value, ending up with a significantly different solution? Given the strong bias revealed by the L23 calculations, the question naturally arises as to whether convergence failure issues could be also be contributing to systematic errors.

\section{Astronomical, atomic, and laboratory data} \label{sec:data}

The astronomical data used for this study is an isolated section (i.e. flanked by continuum, line-free regions) within an extensive quasar absorption system at $z_{abs} = 1.15$ towards the quasar HE\,0515$-$4414 \cite{Reimers1998}. Specifically, our study focuses on the lower redshift side the large complex, which we refer to as the ``left region''. Full details and plots for the left region are described in detail in L23. We use a subsection, rather than the whole absorption system, (a) because the kinematic structure of the left region is such that it provides tight constraints, and (b) to make computing times feasible. Recent studies of this absorption system are given by \cite{Milakovic2021, Murphy2022}, L23, and the astronomical (and all other) data used in the present paper are exactly the same ESPRESSO data as used in these last two studies.

We opted to use the HE\,0515$-$4414 ESPRESSO data for the present study because the analysis carried out in the present paper is directly motivated by the different approaches in \cite{Murphy2022} (hereafter referred to as M22) and L23 that use the exact same data. In the former, as in some previous studies, $\daa=0$ was fixed throughout model development, and only introduced as a free parameter after a completed absorption system model had been derived. In the latter, the same approach was carried out, but with an important addition: as well as fixing $\daa=0$ throughout model development, three other fixed values were also tried, $\daa=-10, +10, +30$ (all in units of $10^{-6}$), for the purpose of demonstrating how the fixed-$\alpha$ approach pushes the solution into a false local minimum. In the present analysis, which is an extension of the L23 study, we examine the impact of changing convergence criteria using the same set of four initially-fixed $\daa$.

All atomic data, transition oscillator strengths $f$, damping coefficients $\Gamma$, laboratory wavelengths including isotope and hyperfine wavelengths, and sensitivity coefficients $q$, are as tabulated in the {\it atom.dat} file provided with version 12.4 of {\sc vpfit}, \cite{web:VPFIT}. All calculations reported in this paper have assumed solar relative isotopic abundances. As pointed out in \cite{Webb1999,Webb2014}, deviations from terrestrial relative isotopic abundances impact significantly on $\daa$. The aim of the present paper is not to examine that issue, which will be addressed in a separate study.

\section{Optimisation method} \label{sec:optimise}

{\sc vpfit} v12.4 is used for all calculations in this paper, providing a few improvements over earlier versions. Nevertheless, the results presented here do not depend on the version used -- we have checked that our findings are replicated using earlier versions.

In \cite{WebbVPFIT2021} we explored different optimisation approaches, applied within the {\sc vpfit} code for modelling high resolution absorption lines or systems. Prior to version 12.3, the code used both Gauss-Newton (GN) and Levenberg-Marquardt (LM) algorithms. The two methods differ in the way in which parameter updates are tuned at each iteration during the optimisation process, for maximum descent in $\chi^2$. In Gauss-Newton optimisation, the vector of optimal parameter updates is given by
\be
\mathbfit{p}_{\mathrm{GN}} = \alpha \mathbfit{p}_{\mathrm{min}} \,, \label{eq:GN}
\ee
where $\gamma$ is a local univariate minimisation parameter whose purpose is to minimise the current value of the merit function $\chi^2$ at each iteration, and where $\boldmath{p}_{\mathrm{min}}$ is the vector of parameter updates prior to tuning via univariate minimisation.

The LM method takes a different approach. Instead of optimising the vector of parameter updates, the Hessian matrix is modified at each iteration using a different univariate minimisation parameter, $\eta$, again to generate the largest reduction in $\chi^2$,
\be
\mathbf{G}_{LM} = \mathbf{G} + \eta \boldsymbol{\mathcal{I}} \,, \label{eq:LM}
\ee
where $\boldsymbol{\mathcal{I}}$ is an identity matrix and $\eta$ is a positive scalar. 
Both GN and LM methods are of course very widely used and both work well. Empirically, in the {\sc vpfit} application, GN sometimes produces the largest $\chi^2$ reduction at some iteration, whilst sometimes LM does. For that reason, {\sc vpfit} (prior to version 12.3) computes both Equations \eqref{eq:GN} and \eqref{eq:LM} and selects, at each iteration, the best option. We refer to that approach as GN-LM. However, in version 12.3, an enhancement was introduced: instead of employing one or the other of Equations \eqref{eq:GN} and \eqref{eq:LM}, {\it both} minimisations are done at every iteration, the obvious goal being a more efficient $\chi^2$ descent. Such a method is evidently neither pure GN nor pure LM, but is a merger of the two approaches, termed ``hybrid optimisation'' (HO) in \cite{WebbVPFIT2021}. 

Another relevant {\sc vpfit} enhancement concerns the way in which the gradient vector and Hessian matrix are calculated; in \cite{WebbVPFIT2021, Lee2021Addendum} we showed how analytic Voigt derivatives avoid problems that occasionally appear using finite difference derivatives and provide better stability in some circumstances. Those changes have been implemented in v12.4.

In all modelling, we imposed an upper limit on the velocity dispersion parameter $b \le 10$ km/s. In all {\sc ai-vpfit} models used in this paper, the SpIC information criterion was used \citep{Webb2021}. Also, we have used a Gaussian instrumental profile for convolution with theoretical Voigt profiles (see L23 for details), as did M22.

\section{Convergence test using compound broadening} \label{sec:convergence}

\subsection{100 AI-VPFIT compound models} \label{sec:100cpmodels}

In L23, we computed a total of 400 left region models: 25 independently derived {\sc ai-vpfit} models for each of 4 initially-fixed $\daa$ values, 2 information criteria (SpIC and AICc), and 2 line broadening mechanisms (turbulent and compound). In the present study, we drop AICc because previous calculations indicate that SpIC is more effective in this context \citep{Webb2021}. We also drop turbulent broadening, as compound is more general, more physically reasonable, and it incorporates turbulent broadening as a limiting case. Our initial set of calculations in the present paper therefore only require 100 models: 25 independently derived {\sc ai-vpfit} models for each of 4 initially-fixed $\daa$ settings, using SpIC and compound line broadening. The 4 initially-fixed $\daa$ settings are $\daa = -10, 0, +10$, and $+30 \times 10^{-6}$ (in both L23 and the present paper). These 100 {\sc ai-vpfit} models provide the starting point for the longer {\sc vpfit} iterations described in Section  \ref{sec:100cpits}.

\subsection{1000 VPFIT iterations for 100 AI-VPFIT compound models} \label{sec:100cpits}

Our goal is to examine whether or not the stopping criteria applied in previous measurements result in convergence. The stopping criterion in {\sc vpfit} is
\be
\frac{\chi^2_n - \chi^2_{n+1}}{\chi^2_n} \le \Delta \label{eq:deltachisq}
\ee
where the subscript $n$ indicates the $n^{th}$ iteration and 
\be
\chi^2 = \sum_{j=1}^M \sum_{i=1}^{N_j} \frac{(d_{i,j} - f_{i,j})^2}{\sigma_{i,j}^2}
\ee
where $N_j$ is the number of data points for the $j^{th}$ transition, the fit being made simultaneously to a total of $M$ transitions. The appropriate choice for $\Delta$ in Eq.\eqref{eq:deltachisq} clearly depends on the data quality and defines the reliability of the result, as well as the total computing time; if $\Delta$ is too large, iterations may terminate too quickly and convergence may not be reached. If $\Delta$ is too small, iterations may proceed indefinitely. In all previous $\alpha$ measurements, a balance between these extremes has been attempted, but without sufficiently rigorous checking, hence the present study. The {\sc vpfit} v12.4 default value of $\Delta$, i.e. the hard-coded value used in the absence of user-definition, is $10^{-3}$. The $\Delta$ values used in the three previous studies of this absorption section are $5 \times 10^{-4}$ \citep{Milakovic2021}, $5 \times 10^{-4}$ (L23); in these both papers, 6 additional iterations were carried out after the stopping criterion had been reached), and $2 \times 10^{-6}$ (M22, without additional iterations).

\begin{figure*}
\centering
\includegraphics[width=0.6\linewidth]{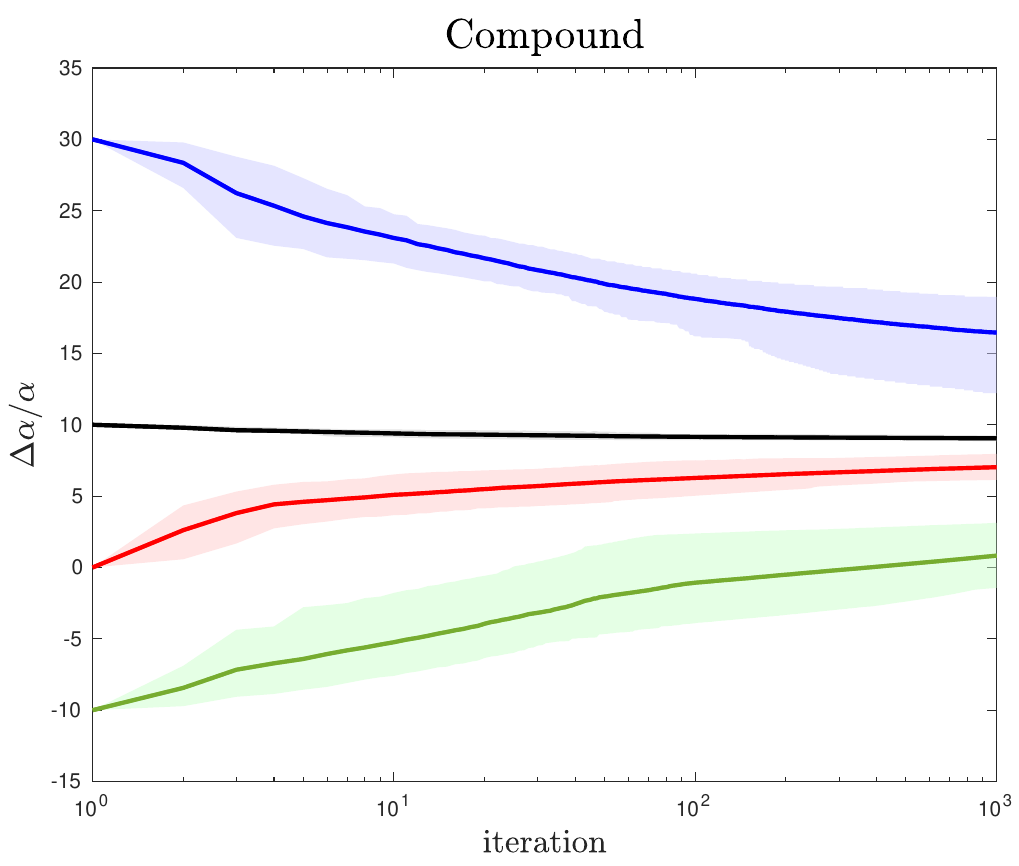}
\caption{Convergence tracks for four sets of 25 initially-fixed $\daa$ {\sc ai-vpfit} models, created using compound broadening. The continuous lines illustrate means over 25 models. Each colour designates a different initially-fixed value of $\daa$ (equal to the value at iteration one). All individual curves are shown in Figs.\ref{fig:daa_details_n} and \ref{fig:daa_details_log}. The shaded areas, which illustrate the impact of model non-uniqueness, encompass empirical $\pm 34$\% ranges. The black ($+10$) values scatter very little so the shaded area is not much larger than the line thickness. $\daa$ is in units of $10^{-6}$. Terrestrial isotopic abundances have been assumed throughout this work. See Section \ref{sec:100cpits} for further details.}
\label{fig:allcp}
\end{figure*}

To explore convergence properties, we {\it remove} the usual requirement defined by Eq.\eqref{eq:deltachisq}, and instead force {\sc vpfit} to iterate 1000 times. Note that whilst the starting models are already ``statistically acceptable'' fits to the data, (i.e. the reduced $\chi^2 \approx 1$), their $\daa$ values are consistently biased towards the starting guesses. Fig.\,\ref{fig:allcp} shows the $\daa$ convergence tracks for initially-fixed $\daa = -10, 0, +10, +30 \times 10^{-6}$. These curves reveal a number of interesting properties:
\begin{enumerate}
\item After a small number of {\sc vpfit} iterations (about 3), each group of 25 models ``spreads out'' to an approximately constant scatter thereafter.
\item Each group, on average, gradually drifts away from the starting $\daa$ value.
\item The $+30, 0$, and $-10$ samples all drift towards the $+10$ model, although on average do not quite reach it, even after 1000 iterations. For those 3 samples, convergence is not reached. For the $+10$ case, drift is very slow i.e. the data appear close to convergence.
\item In all 4 cases studied, the drift rate reduces substantially beyond iteration $\sim$$200-300$. This is more easily seen in the linear plots (Appendix \ref{sec:details}).
\item In 3 of the 4 cases studied, the spread is large compared to the statistical error. But in one case ($+10$), the behaviour is completely  different; both the scatter and the drift are very small and the curves appear stable and consistent.
\end{enumerate}

These results show that initially-fixed $\daa$ creates long flat ``canyons'' in $\chi^2$--$\daa$ space, such that convergence within a reasonable number of iterations is generally unlikely (unless the first guess just happens to coincide with the preferred $\daa$ solution). The curves in Fig.\,\ref{fig:allcp}, whose properties are enumerated above, indicate that the $+10$ model is preferred over the other cases. The properties of the $+10$ models also suggest that if so-called ``blinding'' is avoided, i.e. $\daa$ is a free parameter from the outset of model construction, the convergence failures seen in the $+30, 0,$ and $-10$ cases will be avoided (simply because if $\daa$ is a free parameter throughout model building, it's value will already (in this case) be around the $+10$,  iteration 0 value in Fig.\,\ref{fig:allcp}, and hence will not evolve much subsequently. Whilst these results are derived using one particular absorption system, such that the details would necessarily be different for some other system, there is no reason to expect the more general characteristics found do not apply broadly.

\begin{figure*}
\centering
\includegraphics[width=0.6\linewidth]{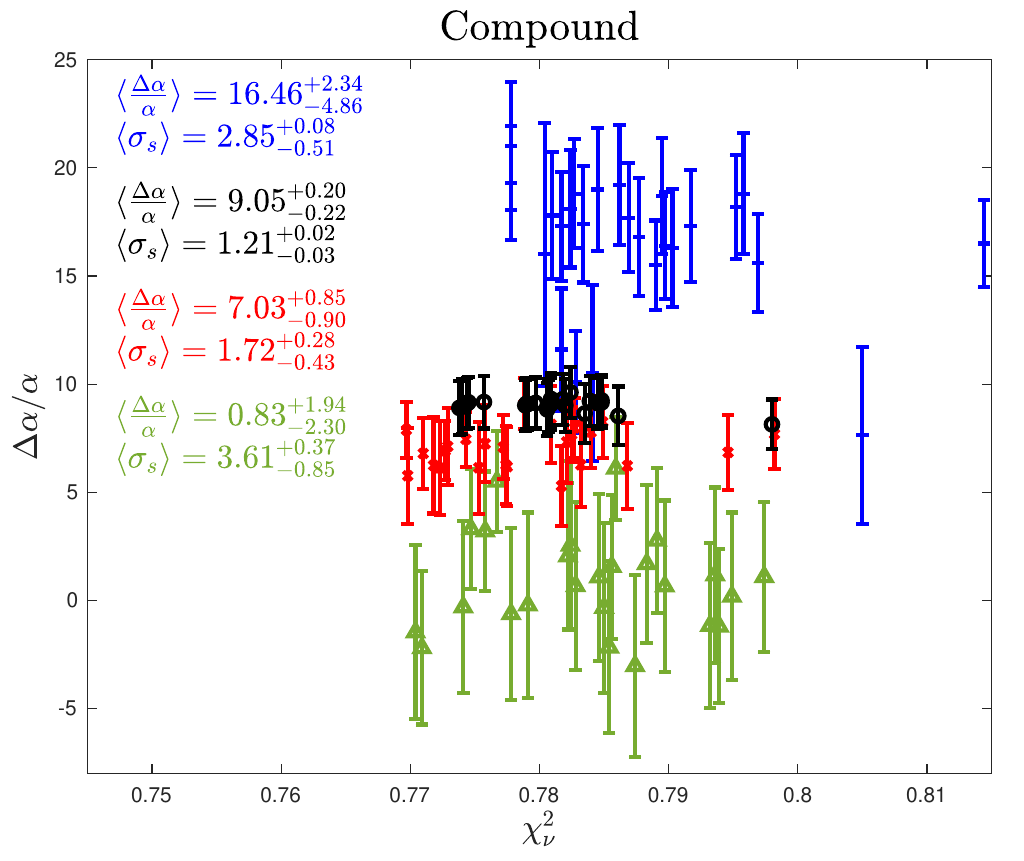}
\caption{$\daa$ measurements for compound broadening models after 1000 iterations. The colour coding (designating four different initially-fixed values of $\daa$) has the same meaning as in Fig.\,\ref{fig:allcp}. $\daa$ is in units of $10^{-6}$. See Section \ref{sec:100cpits}}
\label{fig:stats_cp}
\end{figure*}

In L23, we showed that solving for $\daa$ after being initially-fixed, the standard stopping criterion was met relatively quickly, such that relatively few iterations took place, and hence the ``final'' $\daa$ changed little from its starting point. Fig.\,\ref{fig:stats_cp} shows the corresponding $\daa$ results for all compound broadening models after 1000 iterations. Comparing initial and final values, a slight reduction in the corresponding statistical (covariance matrix) uncertainties is seen, from $(\sigma_{+30}, \sigma_{+10}, \sigma_{0}, \sigma_{-10}) = (3.20, 1.27, 1.75, 3.73) \times 10^{-6}$ in L23, to $(2.85, 1.21, 1.72, 3.61) \times 10^{-6}$ here. This is expected because at the starting point, the $\chi^2$--$\daa$ space is slightly flatter than at the 1000$^{\textrm{th}}$ iteration.

\section{Convergence test using turbulent broadening}

In the M22 study of the $z_{abs} = 1.15$ absorption system towards the quasar HE\,0515$-$4414, as already explained, fixed $\daa=0$ was used throughout the {\sc vpfit} model building process. $\daa$ was only allowed to vary freely once the kinematic structure of the absorption system had been established. Once $\daa$ is introduced as a free parameter, and {\it all} model parameters allowed to vary, the kinematic details evolve a little further, but not much.

\subsection{25 AI-VPFIT models using turbulent broadening} \label{sec:25tbmodels}

As shown in L23, extensive {\sc ai-vpfit} calculations (published after the M22 study) revealed that initially-fixed $\daa$ creates substantial bias. In light of the various results described above, we emulate the M22 analysis, to see if we can reproduce (and hence understand) why that measurement does not reflect the ``correct'' result for that system\footnote{By ``correct'', we mean a statistically unbiased measurement, in this specific case, based on terrestrial isotopic relative abundances, and using turbulent line broadening.}. We therefore carried out an additional set of calculations. For the purposes of illustrating how one can arrive at the M22 $\daa$ measurement, the relevant starting models from L23 are the 25 independent {\sc ai-vpfit} models derived using fixed $\daa=0$ and turbulent line broadening.

\subsection{1000 VPFIT iterations for 25 AI-VPFIT turbulent models} \label{sec:25tbits}

We can thus use the 25 models described above again as a starting point, evolving them further using {\sc vpfit v12.4}, allowing 1000 iterations. None of these 25 models (which involved no human decisions in their construction) would be expected to correspond to the M22 manual process {\it specifically}. However, the {\sc ai-vpfit} process (random placement of trial components, as described in \cite{Lee2020AI-VPFIT}), together with the Monte Carlo generation of 25 independent models, provides a range of models that are representative of those produced by a human subjective process.

\begin{figure*}
\centering
\includegraphics[width=0.6\linewidth]{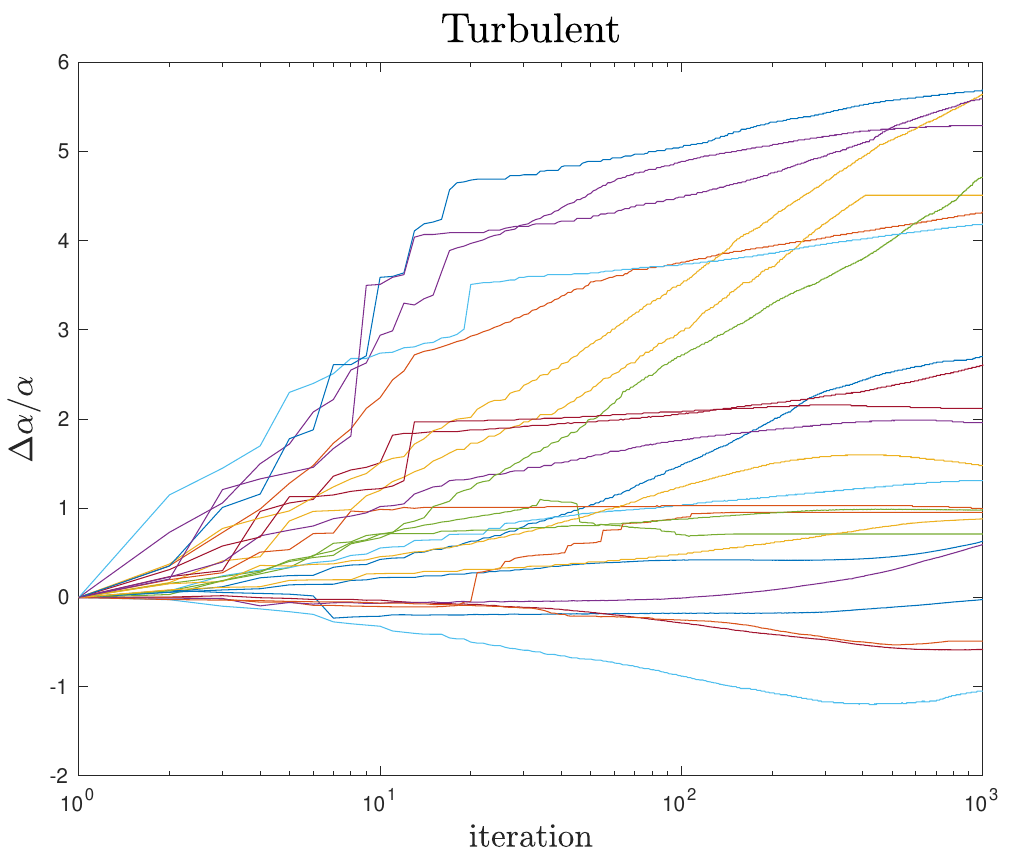}
\caption{Convergence tracks for 25 initially-fixed $\daa=0$ {\sc ai-vpfit} models, created using turbulent broadening. $\daa$ is in units of $10^{-6}$. See Section \ref{sec:25tbits}.}
\label{fig:25tb}
\end{figure*}

Fig.\,\ref{fig:25tb} shows the same general feature as those seen in Fig.\,\ref{fig:allcp}; the models spread in $\daa$ as iterations proceed. After 1000 iterations, the spread is large, and in most cases, convergence does not appear to have been achieved. Interestingly, the compound models appear to reach an approximately constant scatter after only $\sim$3 iterations, but this requires around 25-30 iterations for the turbulent models.

\begin{figure*}
\centering
\includegraphics[width=0.6\linewidth]{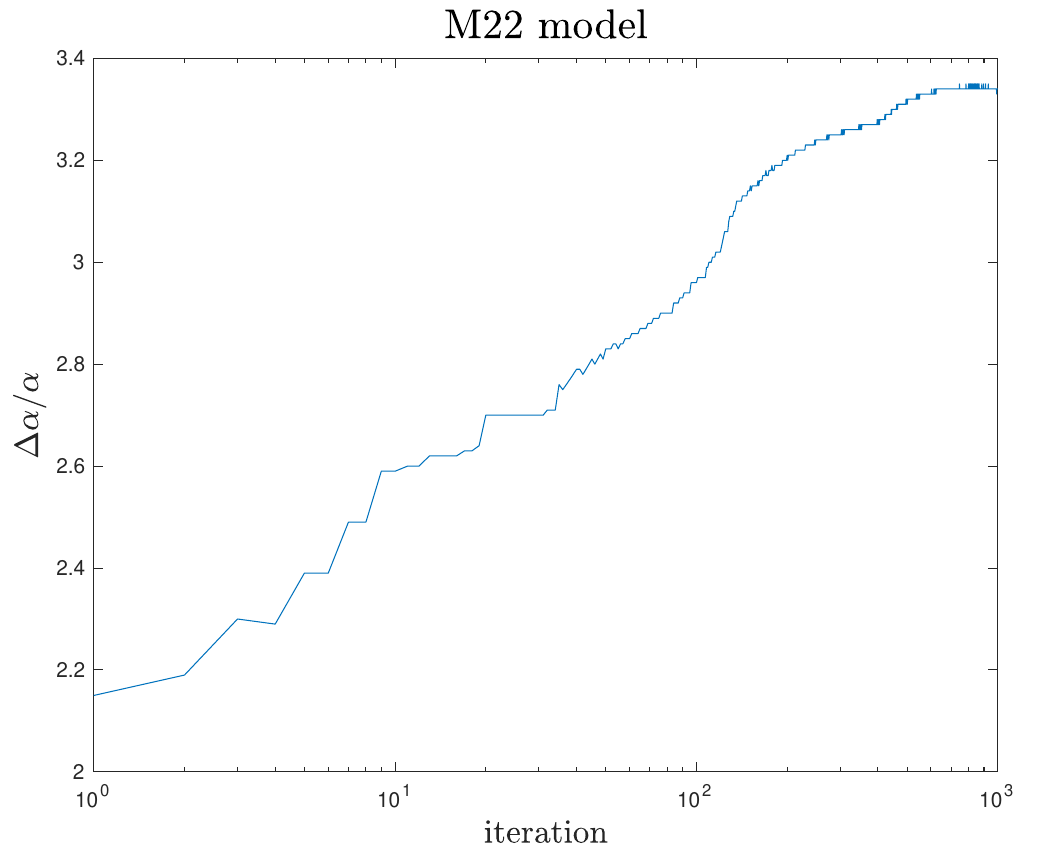}
\caption{Convergence track for the M22 model, created interactively using turbulent broadening and initially-fixed $\daa=0$. The published measurement is the starting point for 1000 iterations (i.e. $\sim$2.2) but this gradually evolves to $\sim$3.3 after 1000 iterations (and may not have reached convergence). $\daa$ is in units of $10^{-6}$. See Section \ref{sec:25tbits}.} 
\label{fig:M22}
\end{figure*}

Fig.\,\ref{fig:25tb} raises an obvious question: did the M22 model converge? M22 argue that any residual convergence uncertainty in their model is small compared to the statistical uncertainty i.e. that, effectively, convergence had been reached. This can easily be explicitly checked by allowing the final M22 model to iterate further. The results of doing this are illustrated in Fig.\,\ref{fig:M22}, which clearly shows that, in fact, the M22 model did {\it not} achieve convergence. The starting point in this case is $\daa=2.15 \times 10^{-6}$. It appears to be that the M22 value of 2.15 was reached after 125 {\sc vpfit} iterations (simply because the $\Delta\chi^2$ stopping criterion used in the M22 study was $10^{-6}$, and it is likely that this criterion was not met, such that {\sc vpfit} is likely to have reached its default maximum of 125 iterations). The result in this case is interesting: as can be seen in Fig.\,\ref{fig:M22}, $\daa$ continuously increases from its starting point, reaching $\sim$3.2 by iteration 1000 (a change corresponding to 36\% of the M22 statistical uncertainty). Fig.\,\ref{fig:25tb} also explains {\it why} the M22 result was obtained; the point (125, 2.2) is typical of the 25 models illustrated.

\begin{figure*}
\centering
\includegraphics[width=0.6\linewidth]{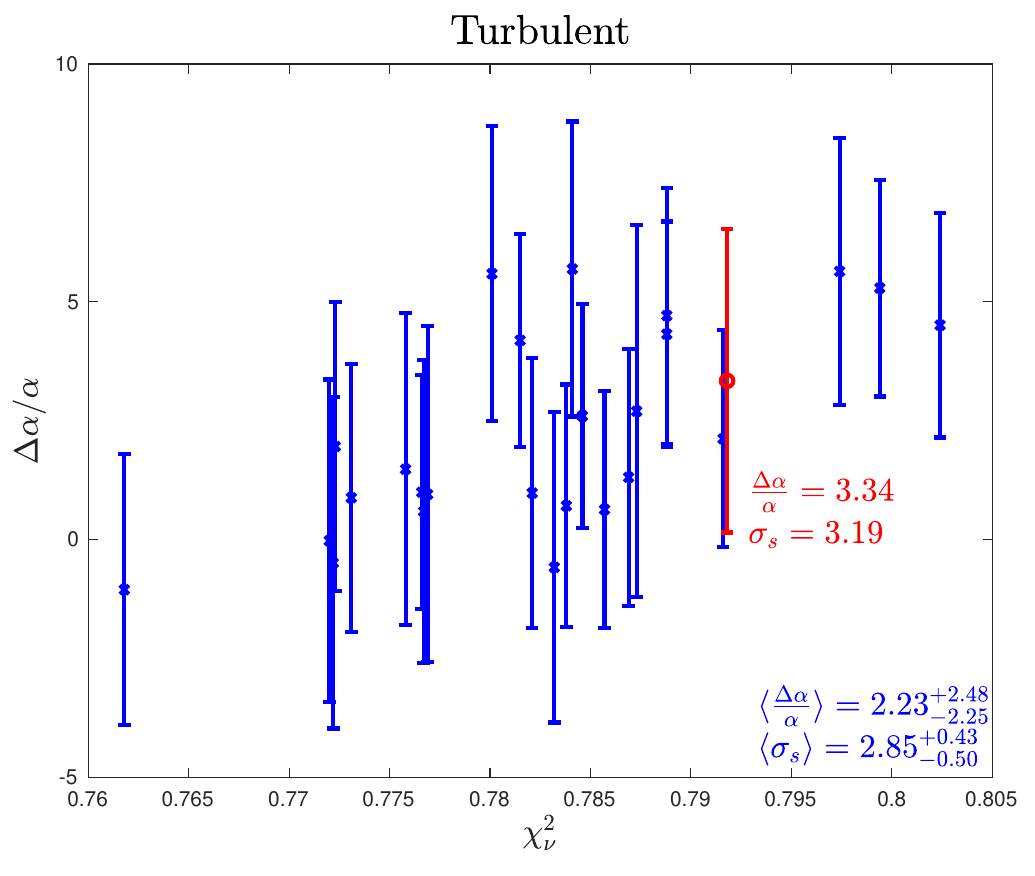}
\caption{$\daa$ values reached after 1000 iterations for all 25 turbulent models (blue points). The single red point is the evolved M22 model. $\daa$ is in units of $10^{-6}$. See Section \ref{sec:25tbits}.}
\label{fig:stats_tb}
\end{figure*}

Fig.\,\ref{fig:stats_tb} shows the $\daa$ values reached after 1000 iterations for all 25 turbulent models (blue points) as well as the single M22 model (red point). This plot again illustrates that the M22 result (both $\daa$ and $\chi^2_{\nu}$) can easily be reproduced using initially-fixed $\daa$ and turbulent broadening. See L23 for a detailed discussion on this point, in particular Table 1 in that paper, which shows that {\sc ai-vpfit} derives models with, on average, close to half the number of absorption components compared to M22 (23.1 instead of 41), with a marginally smaller $\chi^2_{\nu}$.

\section{Summary and implications for fine structure constant measurements}

We have carried out a series of {\sc ai-vpfit} and {\sc vpfit} calculations, to examine convergence properties when measuring $\daa$ in quasar absorption systems. To do so we focused on one particular, very well-studied, system: the left region (as defined in L23) of the $z_{abs} = 1.15$ towards the quasar HE\,0515$-$4414. No previous studies of convergence issues in this context have been done to our knowledge. The specific goal of our study  was to find out how convergence proceeds when ``blinding'' is used, i.e. imposing an initially-fixed $\daa$. Of course, in our conclusions here, we do not criticise blinding methods generically; a comprehensive description of the usefulness of blinding methods used in particle physics in particular is provided in \cite{Harrison2002}. Nevertheless, we have demonstrated the severe bias that may result if great care is not taken with the way in which a blinding method is applied.

The main result of this work is as follows: when models are developed using fixed $\daa$, it is unlikely that subsequently releasing $\daa$ as a free parameter will ever result in convergence, and hence the measurement remains strongly biased towards the fixed value. This kind of ``blinding'' should not be used in this context, and the parameter $\daa$ should be allowed to vary freely in the early stages of model building and left as a free parameter throughout. 

The structure of the specific absorption system we have studied here is complex. It is possible that simpler systems, requiring fewer components, may suffer less from the convergence difficulties we have found. Nonetheless, the results derived here provide a strong warning that substantial bias can occur, the clear implication being that any previous measurement that has been derived using initially-fixed $\daa$ should be reworked.

\section*{Acknowledgements}
{\sc ai-vpfit} calculations were carried out using the OzSTAR supercomputer at the Centre for Astrophysics and Supercomputing at Swinburne University of Technology. This research is based on observations collected at the European Southern Observatory under ESO programme 1102.A-0852 (PI: P. Molaro). We thank that team for making their co-added spectrum publicly available. We also thank Dinko Milakovi\'c for comments on an early draft of this paper.

\section*{Data Availability}
The ESPRESSO spectra and associated files used for this analysis are available at \url{https://doi.org/10.5281/zenodo.5512490}. The {\sc vpfit} and {\sc ai-vpfit}model files are available on request from the authors.

\bibliographystyle{mnras}
\bibliography{convergence2}

\appendix

\section{Further plots} \label{sec:details}

\begin{figure*}
\centering
\includegraphics[width=0.49\linewidth]{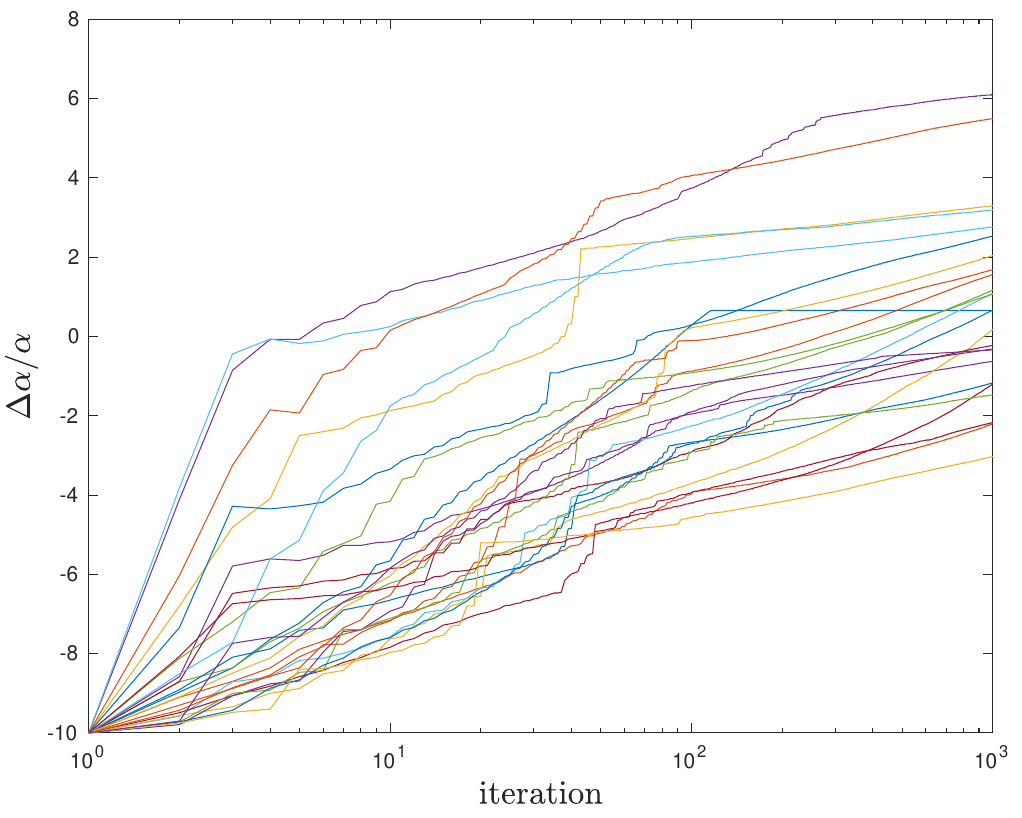}
\includegraphics[width=0.49\linewidth]{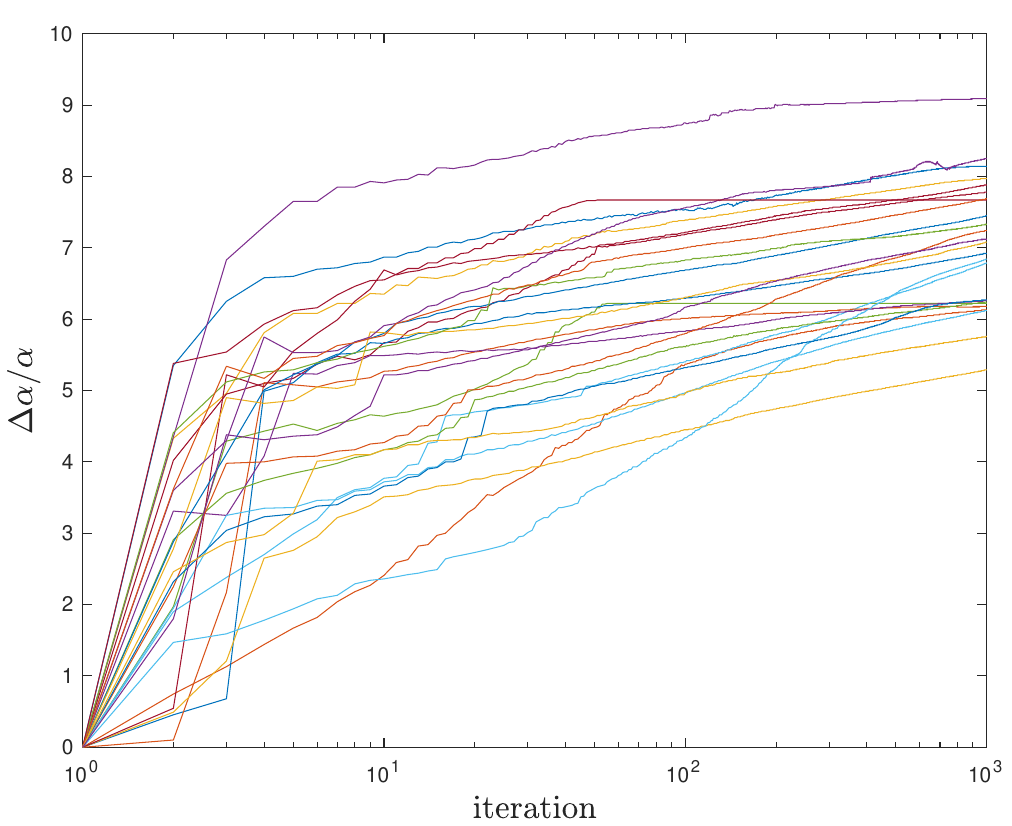}
\includegraphics[width=0.49\linewidth]{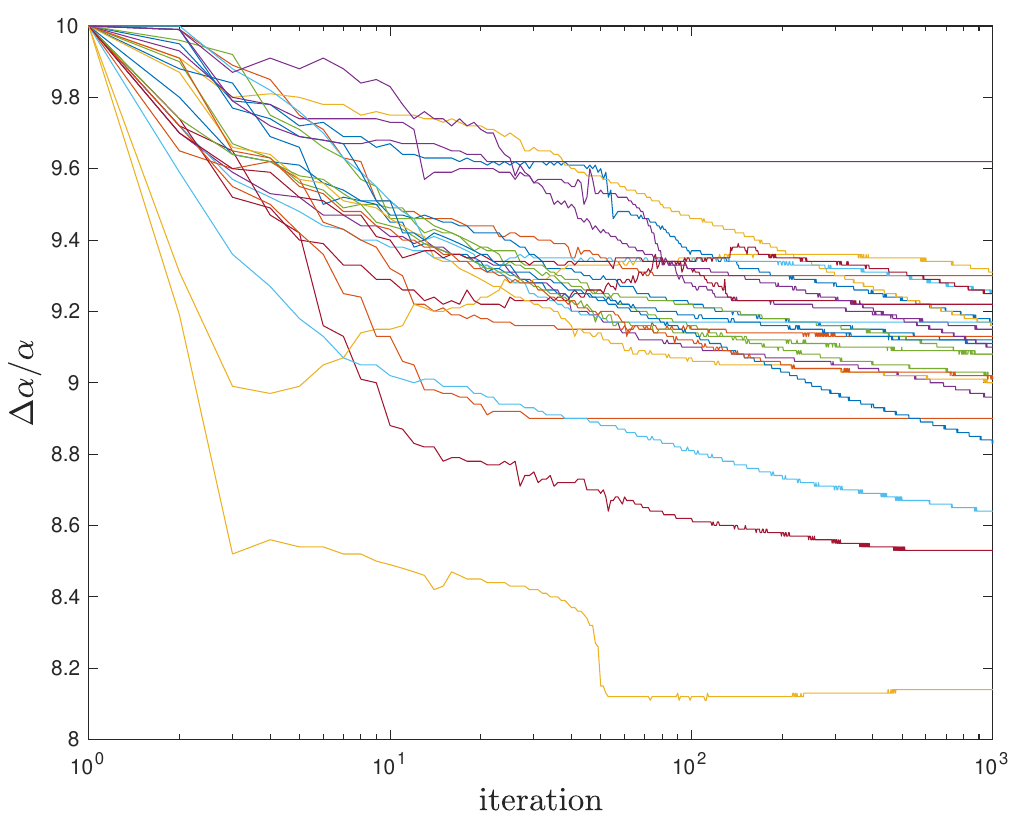}
\includegraphics[width=0.49\linewidth]{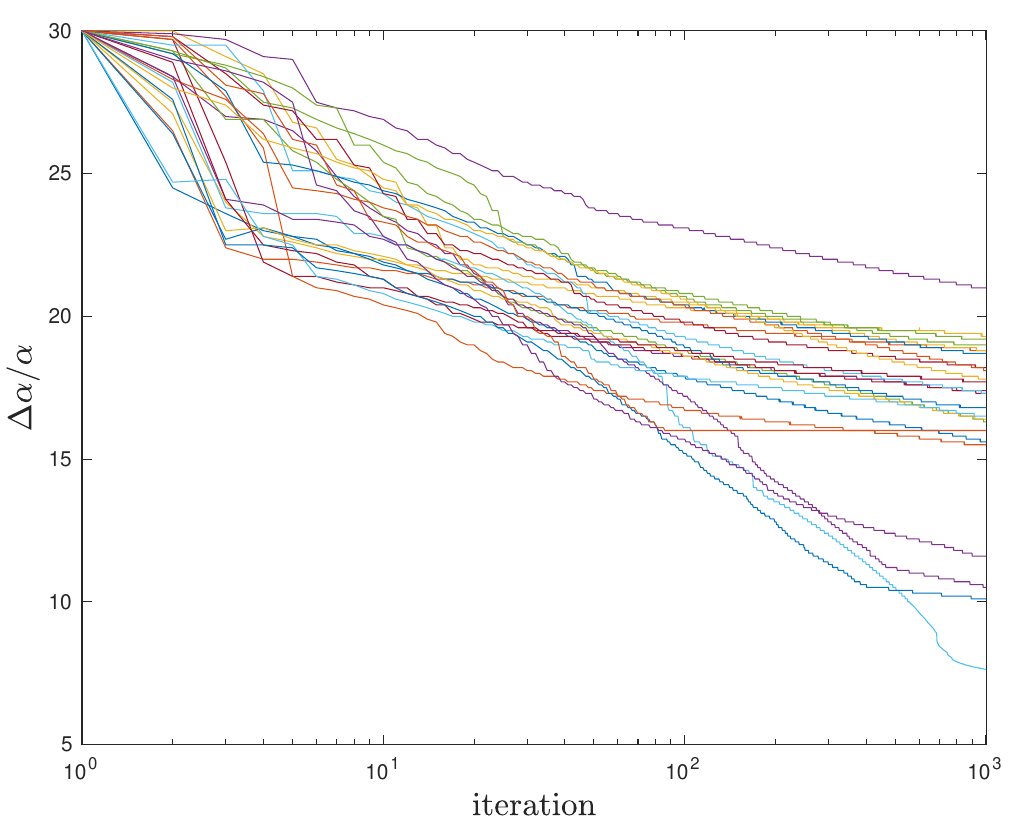}
\caption{Individual convergence tracks for initially-fixed $\daa$ for all four compound broadening cases. $\daa$ is in units of $10^{-6}$.}
\label{fig:daa_details_log}
\end{figure*}

\begin{figure*}
\centering
\includegraphics[width=0.49\linewidth]{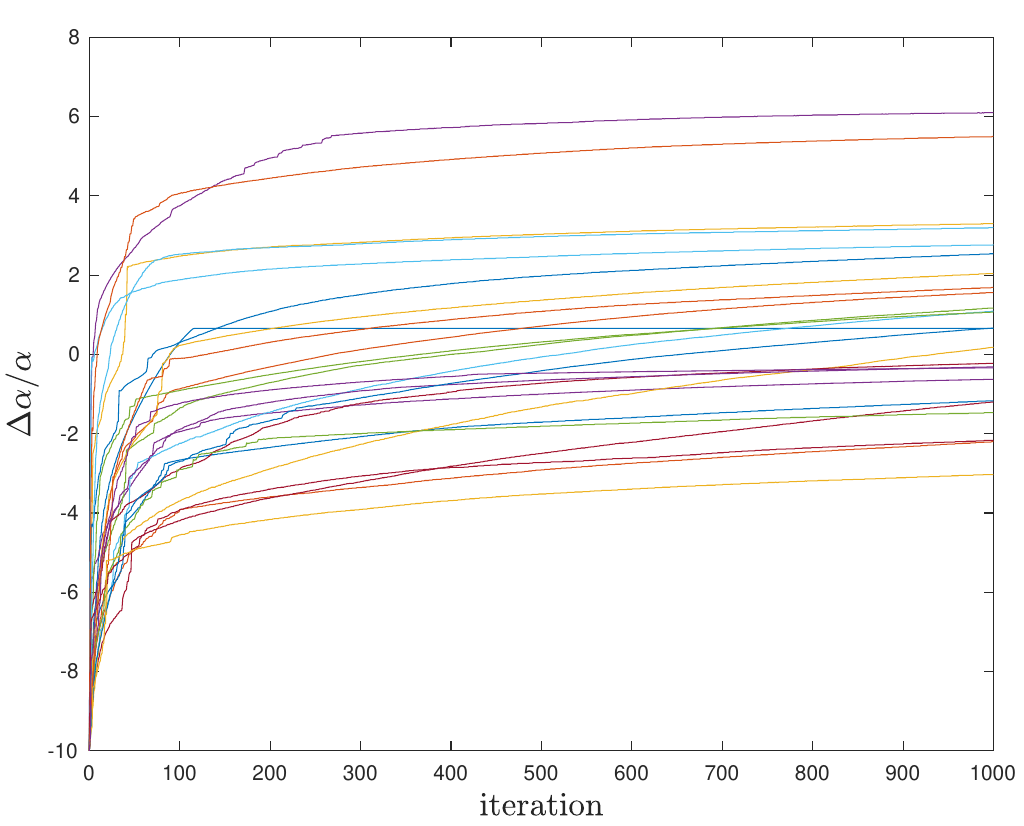}
\includegraphics[width=0.49\linewidth]{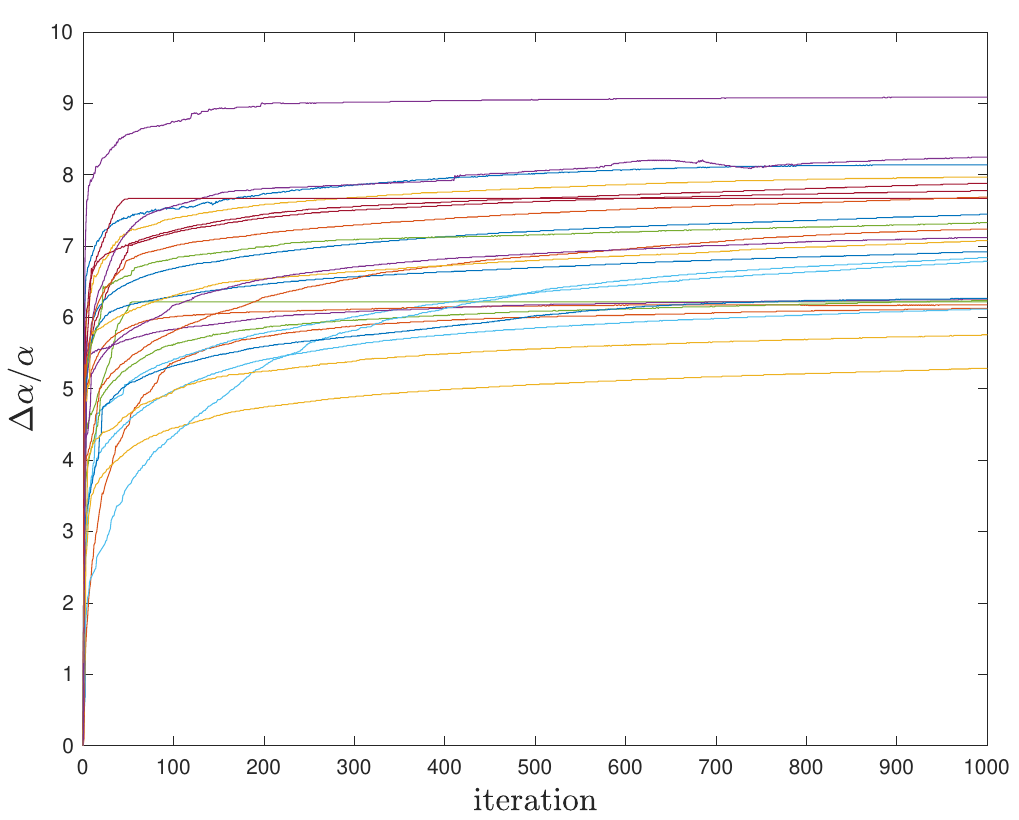}
\includegraphics[width=0.49\linewidth]{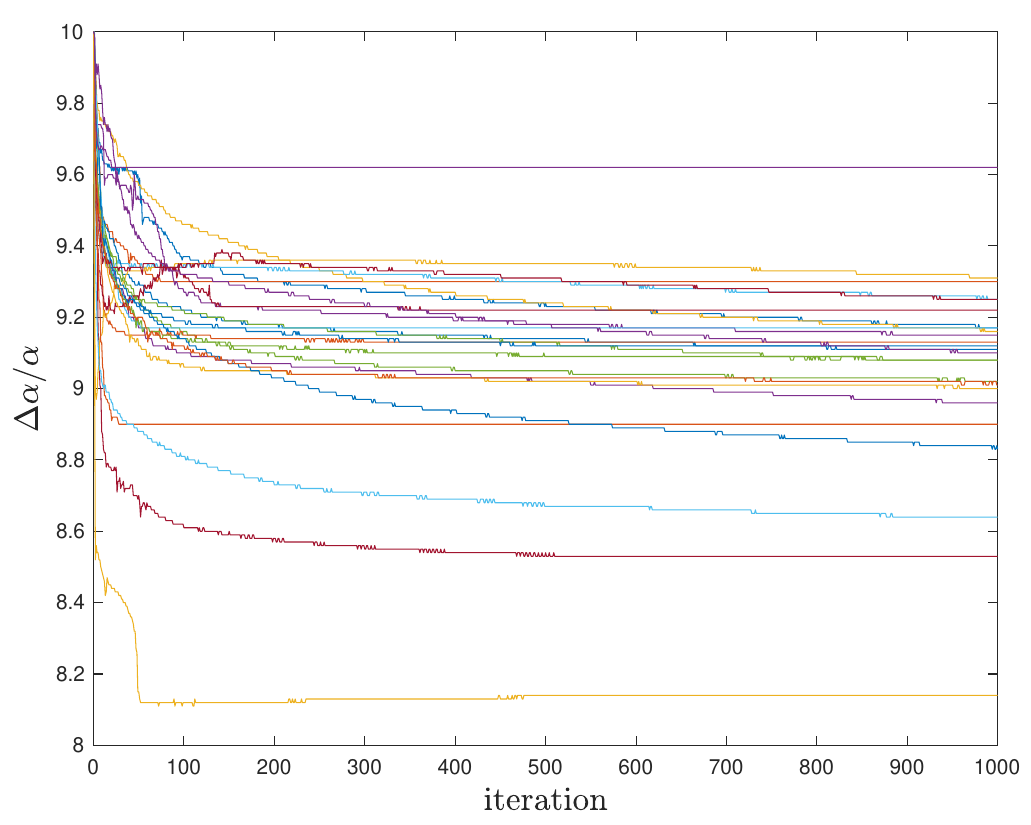}
\includegraphics[width=0.49\linewidth]{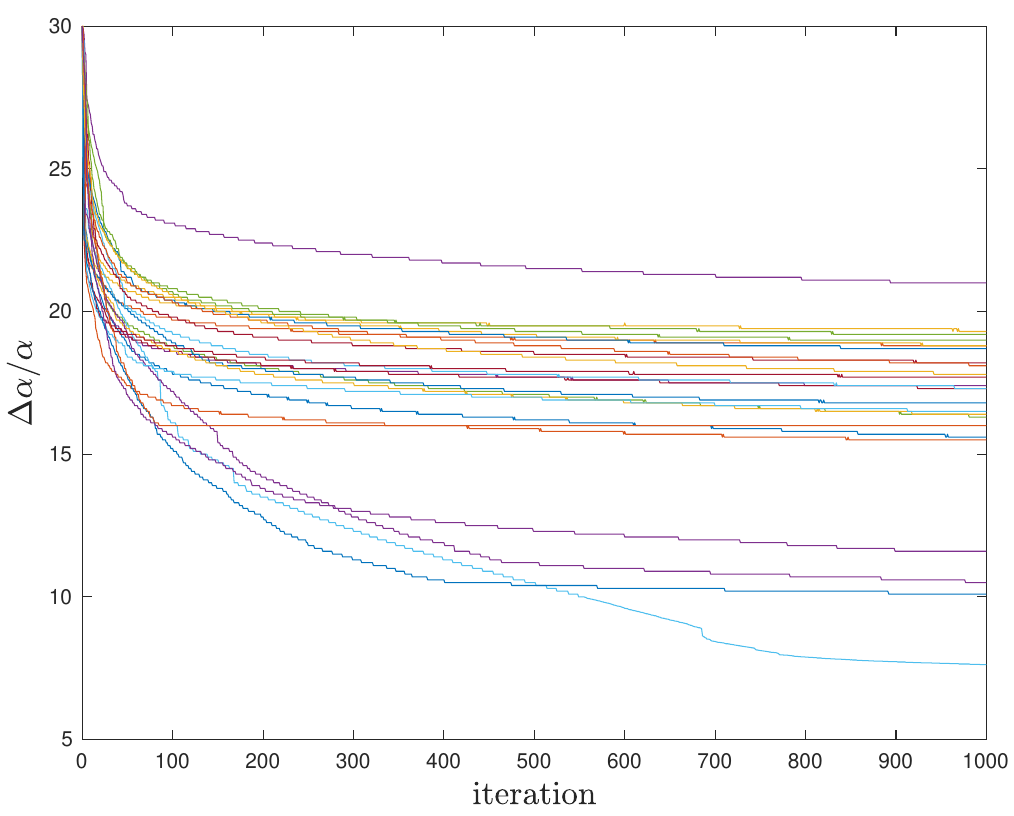}
\caption{Same as Fig.\,\ref{fig:daa_details_log} except abscissa is linear. $\daa$ is in units of $10^{-6}$.}
\label{fig:daa_details_n}
\end{figure*}

\bsp
\label{lastpage}
\end{document}